\documentclass[aps, twocolumn, superscriptaddress]{revtex4-2} 

\usepackage[switch]{lineno}
\usepackage{amsmath}
\usepackage{amssymb}
\usepackage{empheq}
\usepackage[parfill]{parskip}
\usepackage[active]{srcltx}
\usepackage{color}
\usepackage{array}
\usepackage{booktabs}
\usepackage{amsfonts}
\usepackage{dsfont}
\usepackage{graphicx}
\usepackage{appendix}
\usepackage{comment}
\usepackage{hyperref}
\usepackage{lineno}
\usepackage{enumerate}
\usepackage{float}
\usepackage{csquotes}

\begin{document}
\setlength{\parindent}{0.5cm}

\title{Dynamics of pulsating swarmalators on a ring}

\author{Samali Ghosh}
\affiliation{Physics and Applied Mathematics Unit, Indian Statistical Institute, 203 B. T. Road, Kolkata 700108, India}

\author{Kevin O'Keeffe}
\affiliation{Starling Research Institute, Seattle, USA}
	
\author{Gourab Kumar Sar}
\affiliation{Physics and Applied Mathematics Unit, Indian Statistical Institute, 203 B. T. Road, Kolkata 700108, India}
	
\author{Dibakar Ghosh}
\email{dibakar@isical.ac.in}
\affiliation{Physics and Applied Mathematics Unit, Indian Statistical Institute, 203 B. T. Road, Kolkata 700108, India}

\begin{abstract}
\hspace{1 cm}  (Received XX MONTH XX; accepted XX MONTH XX; published XX MONTH XX) \\ \\
We study a simple one-dimensional model of swarmalators, a generalization of phase oscillators that swarm around in space as well as synchronize internal oscillations in time. Previous studies of the model focused on Kuramoto-type couplings, where the phase interactions are governed by phase \textit{differences}. Here we consider Winfree-type coupling, where the interactions are \textit{multiplicative}, determined by the product of a phase response function $R(\theta)$ and phase pulse function $P(\theta)$. This more general interaction (from which the Kuramoto phase differences emerge after averaging) produces rich physics: six long-term modes of organization are found, which we characterize numerically and analytically.

\noindent \\
DOI: XXXXXXX
\end{abstract}
\maketitle

\section{Introduction}
An interplay between synchronization \cite{winfree1980geometry, rosenblum2003synchronization, pecora1990synchronization, boccaletti2002synchronization, pal2024directional} (self-organization in time) and swarming \cite{bialek2012statistical, fetecau2011swarm, darnton2010dynamics, mogilner1999non} (self-coordination in position) arises all over nature and technology, from biological microswimmers \cite{riedel2005self,quillen2021metachronal} and colloidal motors \cite{yan2012linking,bricard2015emergent} to magnetic domain walls \cite{hrabec2018velocity} and robotic swarms \cite{barcis2020sandsbots, talamali2021less, monaco2020cognitive}. Tanaka \textit{et al.} \cite{tanaka2007general, iwasa2010hierarchical, iwasa2010dimensionality} gave one of the first mathematical treatments of this dual kind of pattern formation by introducing a universal model of chemotactic oscillators, oscillators that are pushed around by chemical gradients that, in turn, influence the oscillators' phases. Later O'Keeffe \textit{et al.} \cite{o2017oscillators} introduced a generalized Kuramoto model of swarming oscillators or ``swarmalators" leading to a new wave of work on swarmalation \cite{lizarraga2020synchronization, o2022collective, sar2023pinning, hao2023attractive, kongni2023phase, hong2023swarmalators, ghosh2023antiphase, xu2024collective,kongni2024expected, ghosh2024amplitude,ceron2024reciprocal,blum2024swarmalators,fasciani2024interactive, lizarraga2024order, smith2024swarmalators,sar2025strategy}.

In most -- if not all -- of the studies above, the phase dynamics have `Kuramoto form', by which we mean the phase dynamics $\dot{\theta}_i$ contain a sum over the sinusoid of the phase differences $ N^{-1}\sum_j \sin(\theta_j - \theta_i)$ (usually with some spatial kernel $K(x_j - x_i)$). Yet this type of coupling, which defines the Kuramoto model \cite{kuramoto1984chemical}, is actually an approximation of a more general type of Winfree coupling defined by: $\dot{\theta}_i = \sum_j R(\theta_i) P(\theta_i)$ \cite{winfree1980geometry}. Here, each of the $j$ oscillators fires a pulse defined by the function $P(\theta)$, which is received by the $i$-th oscillator according to $R(\theta)$. This pulsatile coupling is present in many real-world systems, such as fireflies and heart cells, but has not yet been studied by the swarmalator community. This paper fills in this gap. As we show, it gives rise to several new states not seen in swarmalator models with regular Kuramoto-type coupling.

\section{Model}\label{model}
The original swarmalator model concerned particles free to move in the plane \cite{o2017oscillators}. It is, however, rather difficult to analyze
\cite{o2024solvable}, so instead researchers have turned to using the simpler \textit{one dimensional swarmalator model} \cite{o2022collective} where the swarmalators movements are confined to a 1d ring $x_i \in \mathbb{S}^1$. This model is
\begin{align}
     \Dot{x}_i &= v_i+\dfrac{J}{N} \sum_{j=1}^{N} \sin{(x_j-x_i)}\cos{(\theta_j-\theta_i)}, \label{eqa} \\     
    \Dot{\theta}_i &= \omega_i+\dfrac{K}{N} \sum_{j=1}^{N} \sin(\theta_j - \theta_i)\cos{(x_j-x_i)}, 
 \label{eqaa}
\end{align}
where $(v_i, \omega_i)$ are the natural velocities and frequencies and $(J,K)$ are the associated coupling constants. The space dynamics in Eq.~\ref{eqa} describe the phase-dependent aggregation, while the phase dynamics in Eq.~\ref{eqaa} capture position-dependent synchronization. This symmetry makes the model one of the few models of mobile oscillators that is tractable \cite{yoon2022sync, o2025global, o2024stability}. It has enabled exact analyses of swarmalators with distributed coupling \cite{o2022swarmalators,hao2023attractive}, phase frustration \cite{lizarraga2023synchronization}, random pinning \cite{sar2023pinning,sar2024solvable,sar2023swarmalators}, periodic forcing \cite{anwar2025forced}, muilti-body interactions \cite{anwar2024collective}, noise \cite{hong2023swarmalators}, and low range coupling \cite{sar2025effects}.

Here we swap the Kuramoto coupling for Winfree coupling $\sin (\theta_j-\theta_i)\rightarrow R(\theta_i)P(\theta_j)$. We leave the spatial dynamics the same. The model we thus study is
\begin{align}
     \Dot{x}_i &= v_i+\dfrac{J}{N} \sum_{j=1}^{N} \sin{(x_j-x_i)}\cos{(\theta_j-\theta_i)}, \label{eq1} \\     
    \Dot{\theta}_i &= \omega_i+ \dfrac{K}{N} \sum_{j=1}^{N} R(\theta_i) P(\theta_j) \cos{(x_j-x_i)}, 
 \label{eq2}
\end{align}
for $i=1, \cdots, N$, where $N >>1$. For simplicity, we work with swarmalators that have zero frequencies and velocities, i.e., $\omega_i=0 =v_i$ for all $i$. Following previous studies \cite{ariaratnam2001phase}, we make the following choices for the pulse and response functions
\begin{align}
    P(\theta)&=1+\cos \theta, \label{eq3}\\
    R(\theta)&=-\sin \theta. \label{eq4}
\end{align}
Next, we define the order parameters,
\begin{align} \label{eq11}
 R_{x(\theta)} &= r_{x(\theta)} e^{\iota \psi_{x(\theta)}} = \frac{1}{N} \sum_{j=1}^{N} e^{\iota x_j (\theta_j)},\ \iota=\sqrt{-1} \end{align}
where $R_{x(\theta)}$ measures the amount of synchrony among swarmalators' positions and their phases. $r_x$ and $r_\theta$ lie between $0$ and $1$ by definition, indicating the overall synchrony among the swarmalators' positions and their phases, respectively. $\psi_x$ and $\psi_\theta$ are the mean position and phase of the overall population. We also define
\begin{align}\label{eq12}
W_{\pm} &= S_\pm e^{\iota \phi \pm} = \frac{1}{N} \sum_{j=1}^{N} e^{\iota (x_j \pm \theta_j)},
\end{align} 
where $W_{\pm}$ measures the correlation between swarmalators' phases $\theta_j$ and their spatial angle $x_j$. We take the maximum of $S_{\pm}$ and define $S=\max\{S_+,S_-\}$. Another order parameter denoted by $R_{2x(\theta)}$ is defined as, 
\begin{align} \label{eq13}
    R_{2x(\theta)} = r_{2x(\theta)} e^{\iota \psi_{2x(\theta)}} = \frac{1}{N} \sum_{j=1}^{N} e^{2\iota x_j(\theta_j)},
\end{align}
which is beneficial to investigate the antiphase synchrony, where a phase difference of $\pi$ is noticed among the swarmalators' phases. When antiphase synchrony is observed, then $r_{2x(\theta)}=1$ but $r_{x(\theta)}\neq 1$.

These order parameters appear naturally in the equations of motion. Converting the trigonometric functions into complex exponentials, we can rewrite Eqs.~\eqref{eq1} and \eqref{eq2} as,
\begin{align}
    \Dot{x}_i=& \dfrac{J}{2} \bigg[ S_+\sin(\phi_+-(x_i+\theta_i))+S_-\sin(\phi_--(x_i-\theta_i))\bigg], \label{eq5} \\
    \Dot{\theta}_i= &-\dfrac{K}{4}\bigg[ S_+\sin{(\phi_+-(x_i-\theta_i))}-S_+\sin{(\phi_+-(x_i+\theta_i))} \nonumber \\ 
    & +S_-\sin{(\phi_--(x_i-\theta_i))}- S_-\sin{(\phi_--(x_i+\theta_i))} \bigg] \nonumber \\
    &-\dfrac{K}{2} r_x\bigg[\sin{(\psi_x-(x_i-\theta_i))}- \sin{(\psi_x-(x_i+\theta_i))}\bigg]. \label{eq6}
\end{align}
We define $\xi_i=x_i+\theta_i$ and $\eta_i=x_i-\theta_i$ which leads to 
\begin{align}\label{eq7}
    \Dot{\xi}_i=& -\dfrac{K}{2} r_x\big(\sin(\psi_x-\eta_i)-\sin(\psi_x-\xi_i)\big)+ \dfrac{1}{N}\left(\dfrac{J}{2}+\dfrac{K}{4}\right)\nonumber \\
    &\sum_{j=1}^{N}\sin(\xi_j-\xi_i)+\dfrac{1}{N}\left(\dfrac{J}{2}-\dfrac{K}{4}\right)\sum_{j=1}^{N}\sin(\eta_j-\eta_i)\nonumber \\
    &-\dfrac{K}{4N}\sum_{j=1}^{N}\left(\sin(\xi_j-\eta_i)-\sin(\eta_j-\xi_i)\right),
\end{align}
\begin{align}\label{eq8}
    \Dot{\eta}_i &= \dfrac{K}{2}r_x\big(\sin(\psi_x-\eta_i)-\sin(\psi_x-\xi_i)\big)+ \dfrac{1}{N}\left(\dfrac{J}{2}+\dfrac{K}{4}\right)\nonumber \\
    &\sum_{j=1}^{N}\sin(\eta_j-\eta_i)+\dfrac{1}{N}\left(\dfrac{J}{2}-\dfrac{K}{4}\right)\sum_{j=1}^{N}\sin(\xi_j-\xi_i) \nonumber \\
    & +\dfrac{K}{4N}\sum_{j=1}^{N}\left(\sin(\xi_j-\eta_i)-\sin(\eta_j-\xi_i)\right).
\end{align}
Simplifying Eq.~\eqref{eq7} and Eq.~\eqref{eq8}, we get,
\begin{align}\label{eq9}
     \Dot{\xi}_i &=-\dfrac{K}{2} r_x\big(\sin(\psi_x-\eta_i)-\sin(\psi_x-\xi_i)\big)+S_+\left(\dfrac{J}{2}+\dfrac{K}{4}\right)\nonumber \\
    &\sin(\phi_+-\xi_i)+S_-\left(\dfrac{J}{2}-\dfrac{K}{4}\right)\sin(\phi_--\eta_i)\nonumber \\
    &-\dfrac{K}{4}\left(S_+\sin(\phi_+-\eta_i)-S_-\sin(\phi_--\xi_i)\right), 
\end{align}
\begin{align}\label{eq10}
    \Dot{\eta}_i &= \dfrac{K}{2} r_x\big(\sin(\psi_x-\eta_i)-\sin(\psi_x-\xi_i)\big)+S_-\left(\dfrac{J}{2}+\dfrac{K}{4}\right)\nonumber \\
    &\sin(\phi_--\eta_i)+ S_+\left(\dfrac{J}{2}-\dfrac{K}{4}\right)\sin(\phi_+-\xi_i)\nonumber \\
    & +\dfrac{K}{4}\left(S_+\sin(\phi_+-\eta_i)-S_-\sin(\phi_--\xi_i)\right),
\end{align}
which contain $S_{\pm}$, $r_{x(\theta)}$ etc., as claimed.

\section{Results} \label{results}
\subsection{Collective states}
\begin{figure*}
    \centering
    \includegraphics[width=2.2\columnwidth]{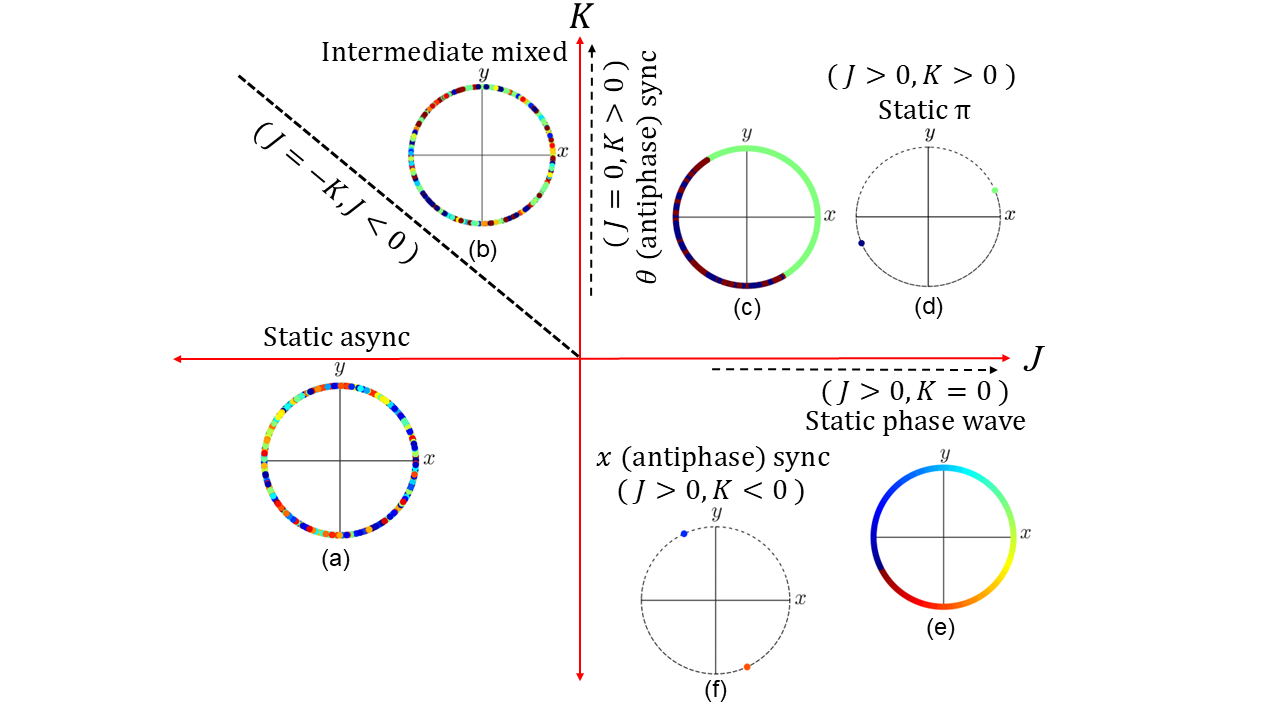}
    \caption{Schematic diagram of the regions for the states for identical swarmalators. (a) Async: $(J, K)=(-0.5,-0.5)$, (b) intermediate mixed state: $(J, K)=(-0.28,0.72)$, (c) $\theta$ antiphase sync: $(J, K)=(0.0,0.5)$, (d) static $\pi$: $(J, K)=(0.5,0.5)$, (e) static phase wave: $(J, K)=(0.5,0.0)$, and (f) $x$ antiphase: $(J, K)=(0.5,-0.5)$. We run the simulations using Heun's method for $N=1000$ number of swarmalators for $T=5000$ time units with step-size $dt=0.1$. Initial positions and phases of the swarmalators are randomly drawn from $[-\pi, \pi]$. All the states are achieved after discarding the first $90\%$ data.}
    \label{fig_1}
\end{figure*}

Next, we explore how our model behaves as the two parameters $(J, K)$ are varied. We drew the phases and positions of $N=10^3$ swarmalators uniformly at random from $[-\pi,\pi]$, then integrated the governing equations using Heun's method. We found the system always settles into six collective states depicted in  Fig.~\ref{fig_1} by varying $J$ and $K$. These collective states are characterized through the order parameters by simultaneously varying $J$ and $K$ as shown in Fig.~\ref{fig2}.

\textit{Static async}: The swarmalators are distributed on the ring in such a way that their positions are static, but their phases evolve asynchronously over time, as shown in Fig.~\ref{fig_1}(a). This state exists for negative $J$ regardless of the value of $K$. Notice Fig.~\ref{fig2}(a)-(d), $S_{\pm}=0$ and $r_{2x(\theta)}=0$ for this state.

\textit{Static $\pi$}: The swarmalators synchronize into two distinct groups, one group is fully sycnhronized with $(x^*,\theta^*)$ and the other is synchronized exactly $\pi$ units away $(x^*+\pi,\theta^*+\pi)$ (see Fig.~\ref{fig_1}(d)). We found that this state exists for $J, K >0$. We observe $S_{\pm}=1$ and $r_{2x(\theta)}=1$ for this state (see Fig.~\ref{fig2}(a)-(d)).

\textit{Static phase wave}: Swarmalators form a phase wave that wraps around the circular domain once $x_i \pm \theta_i + C$ (notice Fig.~\ref{fig_1}(e)). The wave can go both clockwise and counterclockwise. In the coordinates $(\xi_i, \eta_i)$, $\xi_i$ is splayed $\left(\xi_i=\frac{2\pi i}{N}\right)$ and $\eta_i$ is locked $(\eta_i=C)$ or vice versa. The order parameters are $(S_+ , S _-)=(1,0)$ or vice versa, depending on whether the phase gradient of the rainbow is clockwise or anticlockwise, and $r_{2x(\theta)}\sim 0 $ for this state (look Fig.~\ref{fig2}(a)-(d)). This state exists only in the special case where the phase dynamics are turned off, $K=0$ (and  $J>0$).

\textit{$\theta$ antiphase sync}: The swarmalators are uniformly distributed along the run. Despite being evenly distributed in space, they organize their phase by separating into two distinct clusters, maintaining a phase difference of $\pi$. Figure~\ref{fig_1}(c) provides a snapshot of this state at a specific time. This state emerges in the region where $J=0$ for any positive value of $K$. Figure~\ref{fig2}(a)-(d) illustrate the order parameters for this state where, $S_{\pm} \approx 0.64$, $r_{2x}$ is on the order of $10^{-2}$, and $r_{2\theta}\sim 1 $.

\textit{$x$ anti-phase sync}: We observe that the swarmalators positionally form two clusters at $\pi$ difference and their phases are not completely disordered; rather, they are clustered around two points at $\pi$ difference. Figure~\ref{fig_1}(f) best illustrates this state. We observe this state in regions $J>0$ and $K<0$. For this state, $S_{\pm}\approx 0.8\pm 0.05$, $\dfrac{S_+}{S_-}\sim 1$, $r_{2x}=1$, and $r_{2\theta} \approx 0.5$ which are depicted in Fig.~\ref{fig2}(a)-(d).

\textit{Intermediate mixed state}: This state is an intermediate state between the static async state and the $\theta$ antiphase sync state. Here, the swarmalators are randomly distributed in terms of position (Fig.~\ref{fig_1}(b)). However, when examining their internal dynamics, they are neither fully desynchronized nor organized into distinct clusters with a phase difference of $\pi$. Look at Fig.~\ref{fig2}(a)-(d), where $S_\pm=0$, $r_{2x}$ in the order of $10^{-2}$, and $0< r_{2\theta}<1$ in this state.

\subsection{Analysis} 
\subsubsection{Stability of the static $\pi$ state}
We analyze the stability of this state by simplifying around the equilibrium point in $(\xi, \eta)$ space. Here, the swarmalators maintain a phase difference of $\pi$ in both position and phase. Therefore, $r_x=0$. For this, we first calculate the eigenvalues of the Jacobian matrix M, 
\begin{align}\label{eq14}
M=
\begin{bmatrix}
Z_\xi & Z_\eta \\
N_\xi & N_\eta
\end{bmatrix}_{2N\times 2N}
\end{align}
where $(Z_\xi)_{ij} = \dfrac{\partial{\Dot{\xi}_i}}{\partial{\xi_j}}$, $(N_\eta)_{ij} = \dfrac{\partial{\Dot{\eta}_i}}{\partial{\eta_j}}$, $(Z_\eta)_{ij} = \dfrac{\partial{\Dot{\xi}_i}}{\partial{\eta_j}}$, and, $(N_\xi)_{ij} = \dfrac{\partial{\Dot{\eta}_i}}{\partial{\xi_j}}$.
We choose $\psi=0$ without loss of generality.

The equilibrium points are $(\xi_i, \eta_i)= (c_1+\pi, c_2+\pi)$. Hence, we can write,
\begin{align}\label{eq15}
    (Z_\xi)_{ij}= \left(\dfrac{J}{2}+\dfrac{K}{4}\right)\textbf{A}_0
     -\dfrac{K}{4}\cos(c_1-c_2)\textbf{A}_1 , 
\end{align}
\begin{align}\label{eq16}
    (N_\eta)_{ij}= \left(\dfrac{J}{2}+\dfrac{K}{4}\right) \textbf{A}_0
    -\dfrac{K}{4}\cos(c_1-c_2)\textbf{A}_1 , 
\end{align}
\begin{align}\label{eq17}
    (Z_\eta)_{ij}= \left(\dfrac{J}{2}-\dfrac{K}{4}\right)\textbf{A}_0
    +\dfrac{K}{4}\cos(c_1-c_2)\textbf{A}_1 , 
\end{align}
\begin{align}\label{eq18}
    (N_\xi)_{ij}= \left(\dfrac{J}{2}-\dfrac{K}{4}\right)\textbf{A}_0
    +\dfrac{K}{4}\cos(c_1-c_2)\textbf{A}_1 , 
\end{align}
\hspace{0.5 cm} where
\begin{align}\label{eq19}
    \textbf{A}_0=\begin{bmatrix}
        -\frac{N-1}{N} &\frac{1}{N} & ... & \frac{1}{N} \\
        \\
         \frac{1}{N} & -\frac{N-1}{N} & ...&\frac{1}{N} \\
         ... &... &... &... \\
         \frac{1}{N} &...&...& -\frac{N-1}{N}
    \end{bmatrix}_{N\times N},\ 
\end{align}
\hspace{0.5 cm} and
\begin{align}\label{eq20}   
    \textbf{A}_1=\begin{bmatrix}
        \frac{N+1}{N} &\frac{1}{N} & ... & \frac{1}{N} \\
        \\
         \frac{1}{N} & \frac{N+1}{N} & ...&\frac{1}{N} \\
         ... &... &... &... \\
         \frac{1}{N} &...&...& \frac{N+1}{N}
    \end{bmatrix}_{N\times N}. 
\end{align}
For this state, $c_1-c_2=2\theta_i= 0,2\pi, 4\pi$. Note that, $M_{s\pi}$ has a dimension of $2N$ since there are two state variables $(\xi,\theta)$ for each of the swarmalators and dim$(A_0)$=dim$(A_1)=N$ as both of them are the sub-blocks of $M$.
Therefore,  
\begin{align}
M_{s\pi}= 
\begin{bmatrix}
        (\frac{J}{2}+\frac{K}{4})\textbf{A}_0-\frac{K}{4}\textbf{A}_1 &~(\frac{J}{2}-\frac{K}{4})\textbf{A}_0+\frac{K}{4}\textbf{A}_1 \\
        \\
        (\frac{J}{2}-\frac{K}{4})\textbf{A}_0+\frac{K}{4}\textbf{A}_1 &~(\frac{J}{2}+\frac{K}{4})\textbf{A}_0-\frac{K}{4}\textbf{A}_1 
    \end{bmatrix}.
\end{align}

One of the eigenvalues of the matrix $\textbf{A}_0$ is $\hat{\lambda}=0$ due to its rotational symmetry and $(N-1)\hat{\lambda}=-1$. Similarly, for $\textbf{A}_1$, the eigenvalues are $0$ and $1$ with multiplicity $1$ and $(N-1)$.
\par For any symmetric matrix 
$ E=\begin{pmatrix}
        C &D\\
        D &C
    \end{pmatrix}$, $\det(E)=\det(C+D)\det(C-D)$.
    \vspace{0.02 cm}
This implies $\det (M_{s\pi})= \det(J\textbf{A}_0) \det(\frac{K}{2}(\textbf{A}_0-\textbf{A}_1))=\det(J\textbf{A}_0)\det(-K\textbf{I})$.

As a result the eigenvalues of $M_{s\pi}$ become
\begin{align}
\lambda=
\begin{cases}
    0, \hspace{30pt} \text{multiplicity } 1  \\
   -J, \hspace{22pt} \text{multiplicity } N-1\\
   -K, \hspace{19pt} \text{multiplicity } N
\end{cases}
\end{align}
It concludes that the static $\pi$ state will be stable if $J>0$ and $K>0$, which matches our numerics.
\begin{figure}
    \centering
    \includegraphics[width=\columnwidth]{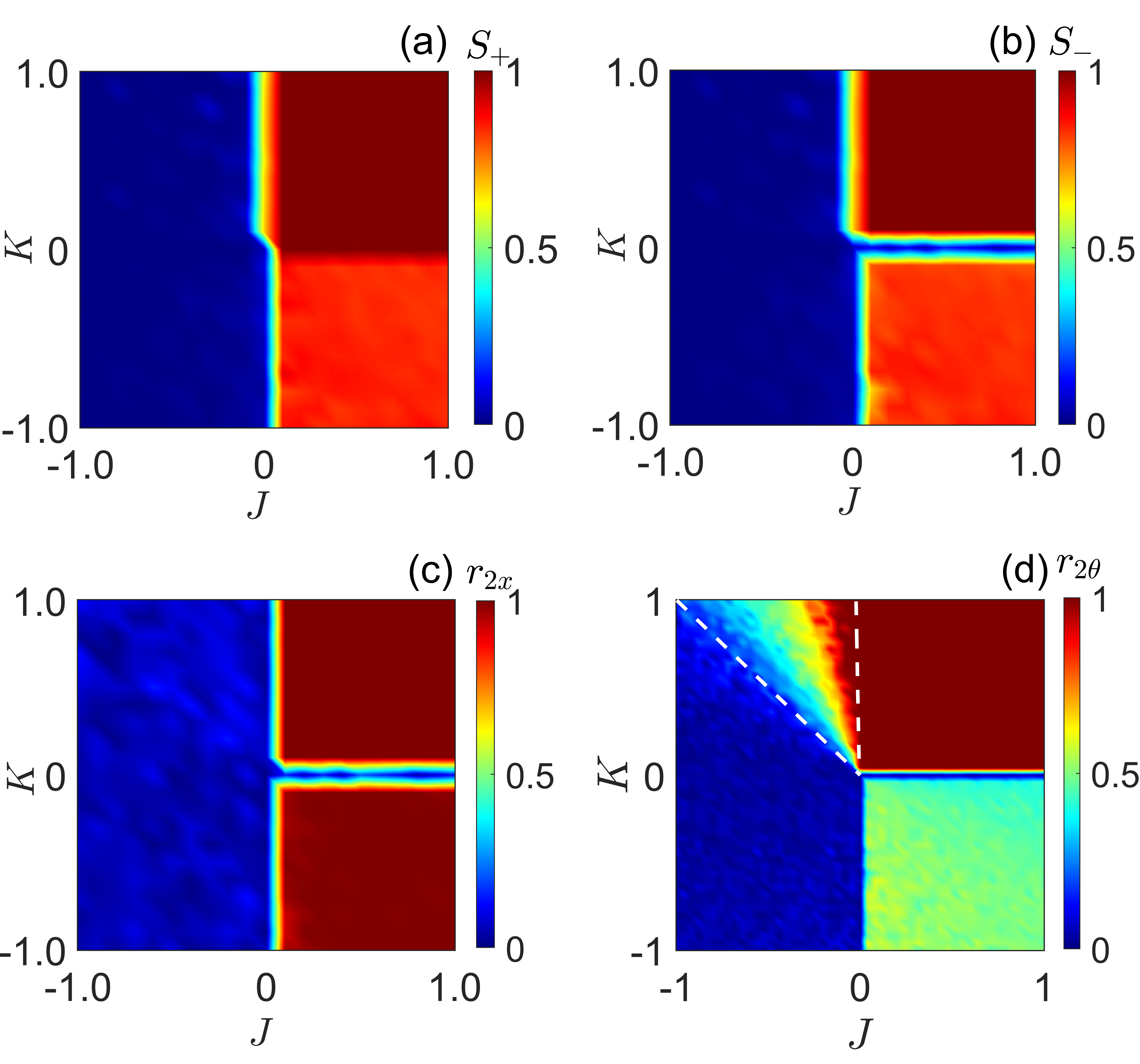}
    \caption{phase space diagram of the order parameters in the $(J, K)$ plane demonstrating the regions of the collective states. (a) $S_+$, (b) $S_-$, (c) $R_{2x}$, and (d)$R_{2\theta}$. We simultaneously vary $J$ and $K$ over the interval $[-1.0, 1.0]$. Simulation parameters are $(dt, T, N)= (0.1, 5000, 10^3)$.[We will provide a detailed explanation of Figure (d) in Section~\ref{stability}].}
    \label{fig2}
\end{figure}

\subsubsection{Stability of the static phase wave state}
The calculation for the stability analysis of this state is the same as before. Firstly, we linearize around the fixed points in the $(\xi,\eta)$ plane and then find the eigenvalues of the Jacobian matrix $M_{spw}$. Here, $x_i = 2\pi i/N = \theta_i$ (for $(S_+,S_-)=(1,0)$ type of phase wave, for $x_i=-\theta_i$, the analysis remains similar). Importantly, this state only exists when $K=0$; it is not a valid fixed point for $K\neq 0$. So we set $K=0$ for our analysis, and seek a stability region defined by the parameter $J$ (numerics suggest this is $J>0$). 

The matrix $M_{spw}$ is written as
\begin{align}
M_{spw} &=\begin{bmatrix}
          Z_\xi & Z_\eta \\
          N_\eta & N_\eta
         \end{bmatrix}_{2N\times 2N} 
         \end{align}
 Now, $C_{ij}= \cos \dfrac{2\pi(i-j)}{N}, S_{ij}= \sin \dfrac{2\pi(i-j)}{N}, \beta_{ij}= C_{ij}+\iota S_{ij}, C_i=\cos\dfrac{2\pi i}{N}, S_i=\sin\dfrac{2\pi i}{N}.$
\begin{align}
\mbox{Therefore,}~~&\dfrac{1}{N}\sum_{\substack{j=1 \\ j\neq i}}^N C_{ij}= \dfrac{1}{N} \sum_{k=1}^{N-1}C_k\nonumber \\
    = &\dfrac{1}{N}\left( \sum_{k=0}^{N-1}C_k-C_0 \right)= -\dfrac{C_0}{N}, 
\end{align}
where $C_0=C_{0,0}=1$.
The elements of the matrix $M_{spw}$ are as follows,
\begin{align}
    (Z_\xi)_{ii} =& -\dfrac{K}{2}r_x\cos \xi_i -\dfrac{1}{N}\left(\dfrac{J}{2}+\dfrac{K}{4}\right)\sum_{j\neq i} \cos (\xi_j-\xi_i) \nonumber \\ 
    &-\dfrac{K}{4N}\Big(\sum_{\substack{j=1 \\ j\neq i}}^N\cos(\eta_j-\xi_i)+\cos (\eta_i-\xi_i)\Big) \nonumber \\
    =&-\dfrac{1}{N}\Big(\dfrac{J}{2}+\dfrac{K}{4}\Big)\sum _j C_{ij}-C_i\Big(\dfrac{K(N+1)}{4N}-\frac{K}{2}r_x\Big)  \nonumber\\
    =&\Big(\dfrac{J}{2}+\dfrac{K}{4}\Big)\dfrac{C_0}{N}-\dfrac{K}{4}\dfrac{(N+1)C_i}{N} -\dfrac{K}{2}r_x C_i ,    \nonumber\\
    (Z_\xi)_{ij} =& \dfrac{1}{N}\Big(\dfrac{J}{2}+\dfrac{K}{4}\Big)\cos(\xi_j-\xi_i)-\dfrac{K}{4N}\cos(\xi_j-\eta_i) \nonumber\\
    =& \Big(\dfrac{J}{2}+\dfrac{K}{4}\Big)\dfrac{C_{ij}}{N}-\dfrac{K}{4}\dfrac{C_i}{N} \nonumber \\
    \mbox{implies,}~~ &Z_\xi= \Big(\dfrac{J}{2}+\dfrac{K}{4}\Big)\textbf{A}_2-\dfrac{K}{4}C_i\textbf{A}_1-\dfrac{K}{2}r_xC_i  \textbf{I}. 
\end{align}

By similar calculation, we obtain the other elements as  
$Z_\eta= (\frac{J}{2}-\frac{K}{4})\textbf{A}_0+\frac{K}{4}C_i A_3+\frac{K}{2}r_x \textbf{I}$, 
$N_\xi=(\frac{J}{2}-\frac{K}{4})\textbf{A}_2+\frac{K}{4}C_i \textbf{A}_1+\frac{K}{2}r_x C_i \textbf{I}$,
and $N_\eta =(\frac{J}{2}+\frac{K}{4})\textbf{A}_0-\frac{K}{4}C_i \textbf{A}_3-\frac{K}{2}r_x \textbf{I}$,
where $A_2= \begin{bmatrix}
        \frac{C_0}{N} &\frac{C_{ij}}{N} &... &\frac{C_{ij}}{N} \\
         \frac{C_{ij}}{N} & \frac{C_0}{N} &...& \frac{C_{ij}}{N} \\
         ... &... &\frac{C_0}{N} &... \\
         \frac{C_{ij}}{N} &...&...& \frac{C_{ij}}{N}
\end{bmatrix}$ and $A_3=\begin{bmatrix}
        \frac{1}{N} &\frac{1}{N} &... &\frac{1}{N} \\
         \frac{1}{N} & \frac{1}{N} &...& \frac{1}{N} \\
         ... &... &... &... \\
         \frac{1}{N} &...&...& \frac{1}{N}
\end{bmatrix}$.

Hence, we can write, \begin{align}
M_{spw}&=\begin{bmatrix}
             J_+\textbf{A}_2&J_-\textbf{A}_0 \\
             J_-\textbf{A}_2&J_+\textbf{A}_0
         \end{bmatrix} 
         +\frac{K}{4}C_i\begin{bmatrix}
             -\textbf{A}_1&\textbf{A}_3 \\
             \textbf{A}_1&-\textbf{A}_3 
         \end{bmatrix} \nonumber \\
         &+\frac{K}{2}r_x \begin{bmatrix}
              -C_i&\textbf{I} \\
               C_i&-\textbf{I}
         \end{bmatrix},
\end{align}
where $J_\pm=\frac{J}{2}\pm \frac{K}{4}$.
The matrix has the same structure as $M_{spw}$ for the regular 1d swarmalator model (with Kuramoto type coupling), whose eigenvalues have been computed exactly \cite{o2022collective}. A direct extension of this analysis provides the eigenvalues for this case. Table~\ref{table1} summarizes these. 
\begin{table}[t!]
    \caption{This table highlights the eigenvalues of the Jacobian and their multiplicities for the static phase wave state for different system sizes.}
    \centering
    \begin{tabular}{ccccc}  \vspace{2pt}
        $N$ value \hspace{1 cm} & (Eigenvalue, Multiplicity)\\ \hline
        \vspace{6pt}
        2 \hspace{1 cm}  &$(0, 4)$\\ 
        \vspace{6pt}
        3 \hspace{1 cm}  &$(0, 4), (-\dfrac{J}{4}, 2)$ \\
        \vspace{6pt}
        4 \hspace{1 cm}  &$(0, 6), (-\dfrac{J}{2}, 2)$ \\
        \vspace{6pt}
        $N\ge 5$ \hspace{1 cm} &$(0, N+1), (-\dfrac{J}{2}, N-3), (-\dfrac{J}{4}, 2)$ \\ \hline
    \end{tabular}
     \label{table1}
\end{table}
The cases $N \leq 4$ are special, while a general pattern holds for $N > 4$. In each case, the stability region is $J>0$, consistent with numerics.

\subsubsection{Stability of the static async state}
We can analyze the stability of this state by going to the continuum limit. Let $\rho(x,\theta,t)$ denote the fraction of swarmalators with positions/phases between $x, x+dx$ and $\theta, \theta+d\theta$ at time $t$. This density obeys the continuity equation
\begin{equation} \label{eq39}
    \Dot{\rho}+\nabla (v \rho) = 0,
\end{equation}
where the velocity $v=(v_x,v_{\theta})$ is given by
\begin{align} \label{eq40}
    v_x &= J \int \sin(x'-x) \cos(\theta'-\theta) \rho(x',\theta',t) \; dx' d\theta', \nonumber \\
    v_{\theta} &= K \int \sin \theta (1+\cos \theta') \cos(x'-x) \rho(x',\theta',t) \; dx' d\theta'.
\end{align}
In the limit $N\rightarrow \infty$, the density in the static async state is $\rho_0=1/(4\pi^2)$~\cite{sar2025effects}. We consider a small perturbation around this state such that
\begin{align} \label{eq41}
    \rho = \rho_0 + \epsilon \gamma(x,\theta,t).
\end{align}
The normalization condition requires $\int \rho(x,\theta, t)dx d\theta= 1$ and therefore we have,
\begin{align} \label{eq42}
    \int \gamma(x,\theta,t) \; dx d\theta = 0.
\end{align} 
The velocity can be written as,
\begin{align} \label{eq43}
    v=v_0+\epsilon v_1 ,
\end{align}
where $v_0$ is the velocity in the static async state, which is zero, and we have $v=\epsilon v_1$, where $v_1(v_1^x, v_1^\theta)$ is the perturbed velocity and is given by, 
\begin{align} \label{eq44}
    v_1^x &= \dfrac{J}{2} \text{Im} \left(W_+^1 e^{-\iota(x+\theta)}+W_-^1e^{-\iota(x-\theta)}\right), \\
    v_1^{\theta} &= -\dfrac{K}{4} \text{Im}\bigg( W_+^1 e^{-\iota(x-\theta)}-W_+^1e^{-\iota(x+\theta)}+W_-^1e^{-\iota(x-\theta)}\nonumber \\
    &- W_-^1e^{-\iota(x+\theta)} + 2 R_x^1 e^{-\iota (x-\theta)}- 2 R_x^1 e^{-\iota(x+\theta)}\bigg). 
\end{align}
and $W_{\pm}^1$ and $R_x^1$ are the perturbed order parameters and are defined as, 
\begin{align} \label{eq45}
    W_{\pm}^1 & = \int e^{\iota(x'\pm \theta')} \gamma(x',\theta') \; dx' d\theta'.
\end{align}
Plugging Eq.~\eqref{eq41} and Eq.~\eqref{eq43} in Eq.~\eqref{eq39} obtains
\begin{align} \label{eq46}
    \Dot{\gamma} + \rho_0 (\nabla v_1) = 0,
\end{align}
where $\nabla v_1= \partial_x v_1^x+\partial_\theta v_1^\theta$.

We can calculate $\nabla v_1$ by determining $\partial_x v_1^x$ and $\partial_\theta v_1^\theta$ as,
\begin{align}\label{eq47}
    \partial_x v_1^x &= -\dfrac{J}{2} \text{Re} \left[W_+^1 e^{-\iota(x+\theta)} + W_-^1 e^{-\iota(x-\theta)}\right], \nonumber \\ 
    \partial_x v_1^\theta&= -\dfrac{K}{4} \text{Re} \Big[W_+^1 e^{-\iota(x-\theta)}+W_+^1e^{-\iota(x+\theta)}+W_-^1e^{-\iota(x-\theta)}\nonumber \\ 
    &+W_-^1e^{-\iota(x+\theta)}+2R_x^1 e^{-\iota(x-\theta)}+2R_x^1e^{-\iota(x+\theta)}\Big]. 
\end{align}
Therefore, we obtain the time evolution of $\gamma$ as,
\begin{align}\label{eq48}
    \Dot{\gamma}(x,\theta,t)&=\dfrac{(2J+K)}{16\pi^2} \text{Re}\Big[W_+^1 e^{-\iota(x+\theta)}+W_-^1 e^{-\iota(x-\theta)}\Big]\nonumber \\
    &+\dfrac{K}{16\pi^2}\text{Re} \Big[W_+^1 e^{-\iota(x-\theta)}+W_-^1 e^{-\iota(x+\theta)}\nonumber \\
    &+2R_x^1 e^{-\iota(x-\theta)}+2R_x^1e^{-\iota(x+\theta)} \Big].
\end{align}

We expand $\gamma(x,\theta,t)$ in terms of complex exponentials~\cite{o2022collective} as the following,
\begin{align}
    &\gamma(x,\theta,t) = \frac{1}{2\pi}\bigg( \alpha_{0,0}(t) + \alpha_{1,0}(t) e^{\iota x}+ \alpha_{0,1}(t) e^{\iota \theta} \nonumber \\
    &+ \sum_{n=1}^{\infty} \sum_{m=1}^{\infty} \Tilde{\alpha}_{n,m}(t) e^{\iota (n x + m \theta)} + \Tilde{\beta}_{n,m}(t) e^{\iota (n x - m \theta)} + \text{c.c.}\bigg),
\end{align}

where c.c. denotes complex conjugate. The order parameters are $W_+^1 = \alpha_{1,1}(t)$, $W_-^1 = \beta_{1,1}(t)$, and $R_x^1 = \alpha_{1,0}(t)$. Moreover, due to the normalization condition \ref{eq42}, we have $\alpha_{0,0}=0$.
By projecting onto the Fourier basis, we extract the following evolution equations
\begin{align}
    \Dot{W_{\pm}^1} &= \frac{J}{8\pi} W_{\pm}^1 + \frac{K}{16 \pi}(W_+^1+W_-^1 + 2 R_x^1), \\
    \Dot{R}_x^1 &= 0 .
\end{align}
This is a (linear) three-dimensional dynamic system in the coordinates $W_+^1, W_-^1$, and $R_x^1$. Therefore, the Jacobian is
\[ M_{async}=
\begin{pmatrix}
\dfrac{J}{8\pi}+\dfrac{K}{16\pi} & \dfrac{K}{16\pi} & \dfrac{K}{8\pi} \\[6pt]
\dfrac{K}{16\pi} & \dfrac{J}{8\pi}+\dfrac{K}{16\pi} & \dfrac{K}{8\pi} \\[6pt]
0  & 0  & 0
\end{pmatrix}.
\]
The eigenvalues of $M_{async}$ are $\lambda = 0,\ \dfrac{J}{8\pi}\ \mbox{and}\ \dfrac{J+K}{8\pi}$. These imply a stability region of
\begin{align}
    & J < 0 , \nonumber \\
    & K < -J
\end{align}
consistent with numerics. We draw this stability threshold in Fig.~\ref{fig2}(d)  by the white dashed lines.

\subsubsection{$x$ (antiphase) sync state}

\begin{figure}[hpt]
    \centering
    \includegraphics[width=1.05\columnwidth]{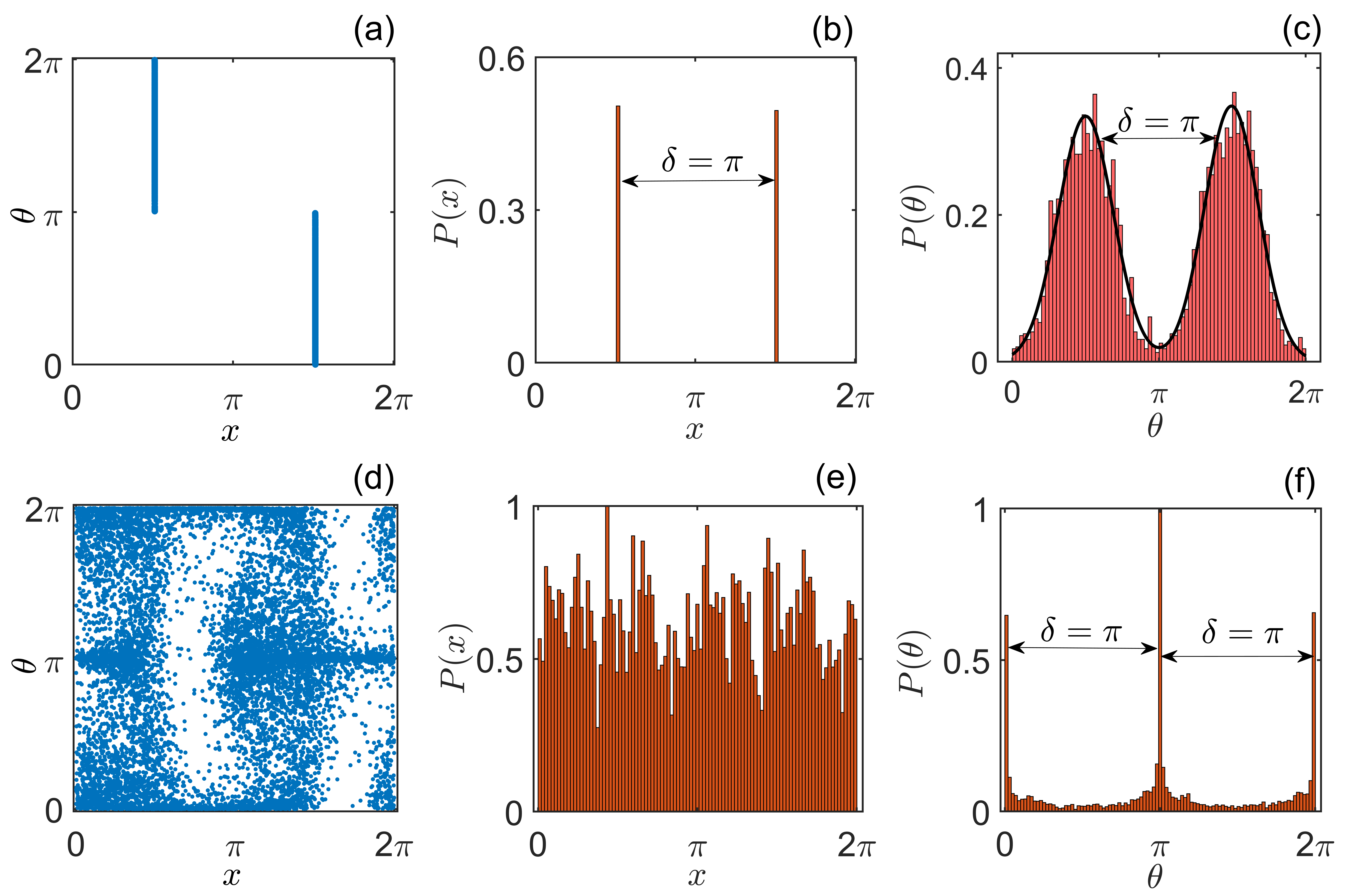}
    \caption{(a) Scatter plot of the $x$ (antiphase) sync state $[(J,K)=(0.5,-0.5)]$ in $(x,\theta)$ plane, (b) \& (c) denote the distribution of $x$ and $\theta$ in the $x$ (antiphase) sync state. (d) Scatter plot of the intermediate mixed state $[(J,K)=(-0.28,0.76)]$ in $(x,\theta)$ plane, (e) \& (f) depict the corresponding distribution of $x$ and $\theta$. The simulation parameters are $(dt,T,N)=(0.1,5000,10^4)$.} 
    \label{fig3}
\end{figure}

This state is quite difficult to analyze. Look at the scatter plots of the swarmalators in Fig.~\ref{fig3}(a). The spatial dynamics are straightforward: there are two equilibrium points shifted $\pi$ units apart, $x_i = C$ and $C+\pi$ (as evident from the position distribution shown in Fig.~\ref{fig3}(b)). We can set the constant $C=0$ with a change of frame. But the phase equilibrium points are hard to identify. Figure~\ref{fig3}(c) shows that the $\theta_i^*$ are distributed in two bell-shaped curves, with populations $p,q:=1-p$, respectively. In general, $p\neq q$, the exact values vary from run to run.

To try and find these equilibrium points, we write the phase equation (Eq.~\ref{eq2}) in the continuum limit, 
\begin{align}
    v_{\theta} &= - \sin \theta \int (1 + \cos \theta^\prime) \cos(x^\prime - x) \rho(x^\prime, \theta^\prime) dx^\prime d \theta^\prime,
\end{align}
where the density of the state is $\rho(x,\theta) = p \delta(x) P_p(\theta) + q \delta(x-\pi)P_q(\theta)$. Here $P_{p,q}(\theta)$ denotes the densities of the two sub-populations. In this notation, the phase velocity becomes
\begin{align}
    v_{\theta}=& - \sin \theta \Big[  p \cos x \int_{\pi}^{2\pi} (1 + \cos \theta^\prime) P_p (\theta^\prime)  d \theta^\prime   \nonumber \\
    &+ q  \cos(\pi-x) \int_{0}^{\pi} (1 + \cos \theta^\prime) P_q(\theta^\prime) d \theta^\prime \Big], 
\end{align}
which implies
\begin{align}
 v_\theta =-\sin \theta \cos x \Big[ p (1 + I_p) - q (1+I_q) \Big],
\end{align}
where $I_{p,q}= \int P_{p,q}(\theta) d \theta$ denote the definite integrals of $P_{p,q}(\theta)$ over their domain. Now, we know $v_{\theta} = 0$ at $x=0$ and $\pi$, which means the term within the brackets must be zero, i.e.,
\begin{align}
p(1+I_p)-q(1+I_q)= 0, \label{tt1}
\end{align}

\begin{figure}[hpt]
    \centering
    \includegraphics[width=\columnwidth]{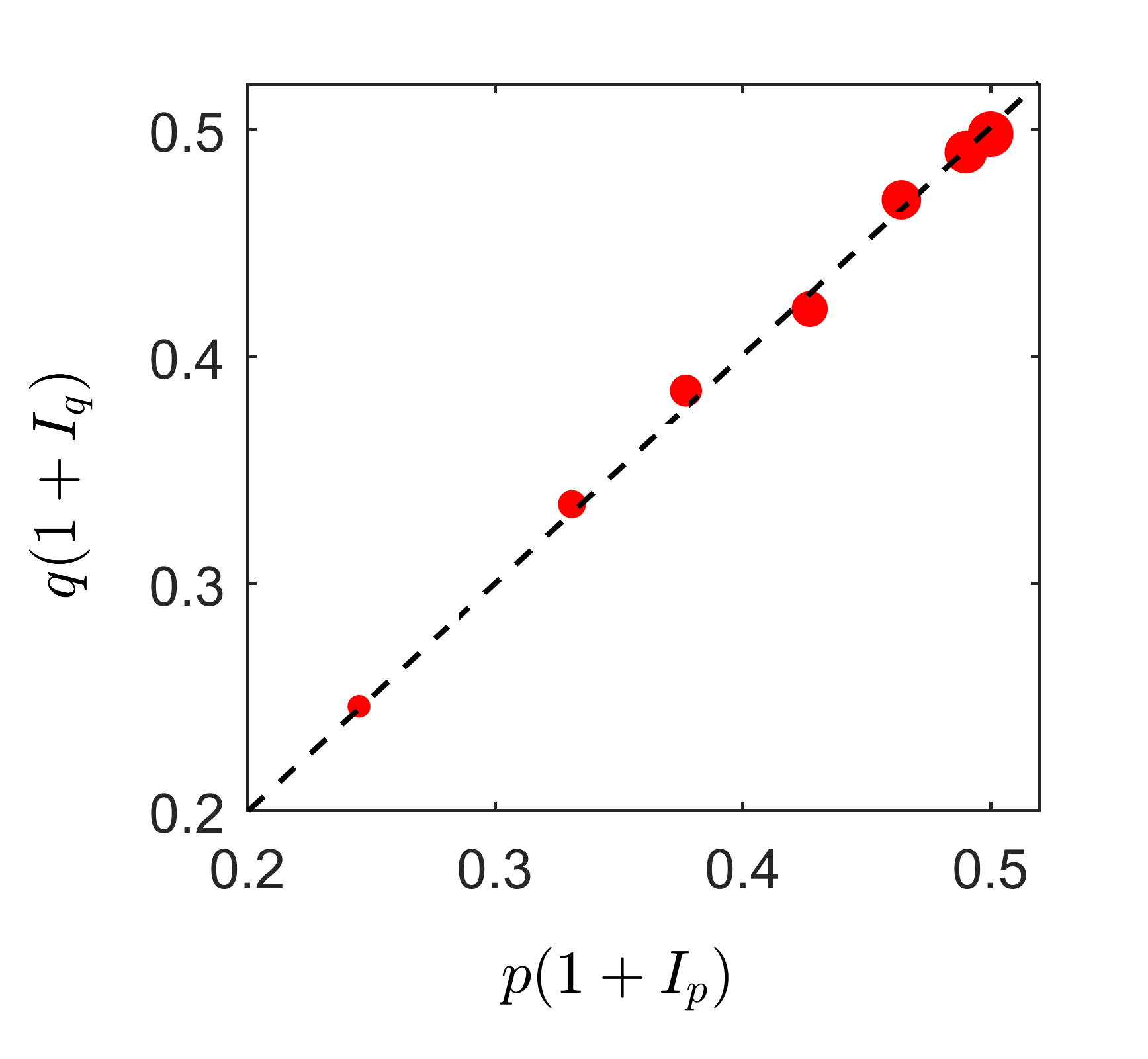}
    \caption{The black dashed line represents the prediction of Eq.~\eqref{tt1}; For numerical validation, we simulate the system for $N=2,3,4,5,20,50,100$ number of particles (as indicated by increasing the size of the red dots), which closely align with our analytical results. Simulation parameters $(dt,T,J,K)=(0.1,5\times10^4,0.5,-0.5)$. After wiping out the transients, we save the last $10\%$ data by taking $100$ realizations.}
    \label{xantiphase_numerics}
\end{figure}

where we have requoted the definition of $I_{p,q}$ for convenience. Eq.~\eqref{tt1} is the equilibrium point condition for this state. Any phase density $P_{p,q}(\theta)$ that satisfies the above is a solution. We have confirmed Eq.~\eqref{tt1} numerically by calculating $(p,q), (I_p, I_q)$ for different population sizes $N$, depicted in Fig.~\ref{xantiphase_numerics}.

Now, getting back to an analytic perspective, what are the equilibrium densities $P_{p,q}(\theta)$ that satisfy Eq.~\eqref{tt1}? There is an infinite number. Yet numerics suggest only one of these is stable, the bell-shaped distributions in Figure~\ref{fig3}(c), which are well fit by a wrapped Gaussian (thick black lines in panel (c)). We are unable to make any headway on proving this analytically. The formal approach would be to linearize around the equilibrium density
\begin{align}
    \rho_0(x,\theta) = p \delta(x) P_p(\theta) + q \delta(x-\pi)P_q(\theta) 
\end{align}
with $P_{p,q}(\theta)$ wrapped Gaussians, in the same way we linearized around the async density $\rho_0 = (4\pi^2)^{-1}$. However, the presence of $\delta(x)$ functions and the wrapped Gaussian makes the calculation significantly more complex and beyond the scope of this paper.

\subsubsection{Intermediate mixed state}\label{stability}
\begin{figure}
    \centering
    \includegraphics[width=\linewidth]{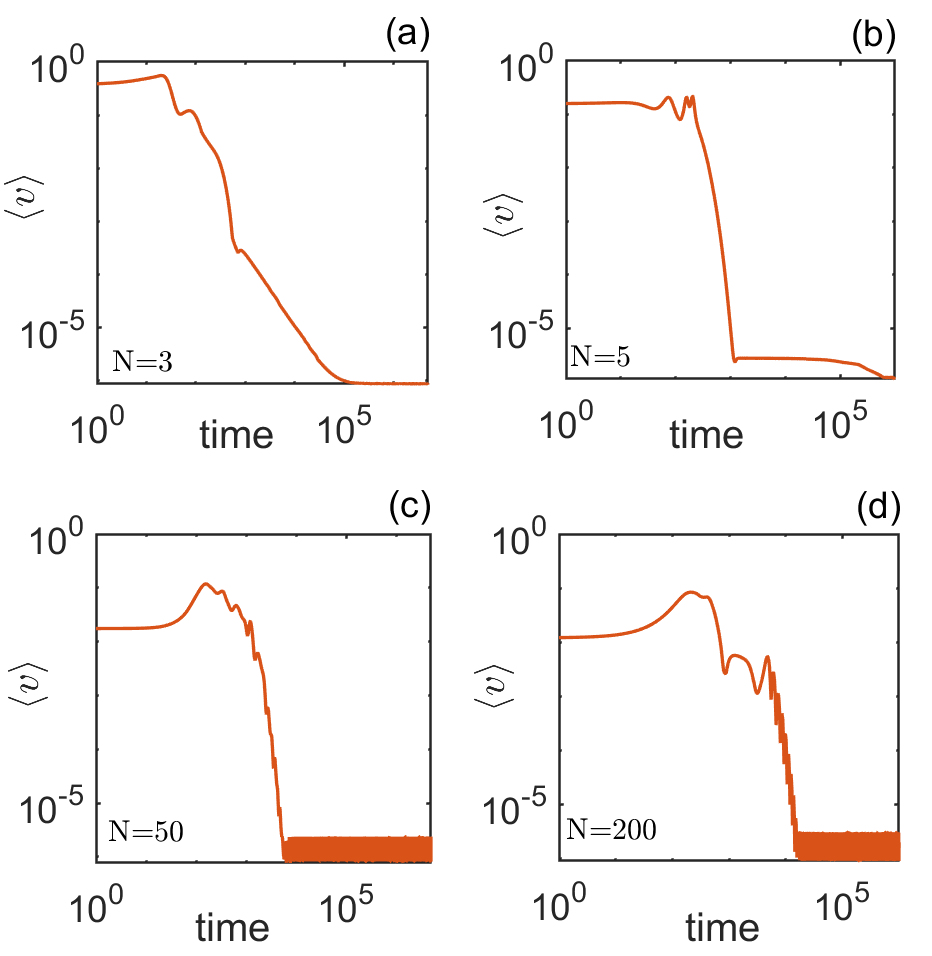}
    \caption{A log-log plot of the time evolution of mean velocity $\langle v \rangle$ of the swarmalators at different values of $N$ for intermediate mixed state. We set $(J,K)=(-0.28,0.76)$.}
    \label{mixed state velocity}
\end{figure}
This state appears in the range between $J=0$ and $J=-K~\forall\ J<0$ in Fig.~\ref{fig2}(d). Figure~\ref{fig3}(d) depicts the scatter plot of the swarmalators in the $(x,\theta)$ plane. Here, the particles are positionally asynchronous. However, a detailed examination of their phases reveals that while some form clusters with a phase difference of approximately $\pi$, the others remain randomly distributed (notice the distribution of positions and phases in Fig.~\ref{fig3}(e) and (f) respectively). We term this the intermediate mixed state, where swarmalators are neither fully antiphase nor entirely asynchronous.

\begin{figure}
    \centering
    \includegraphics[width=\linewidth]{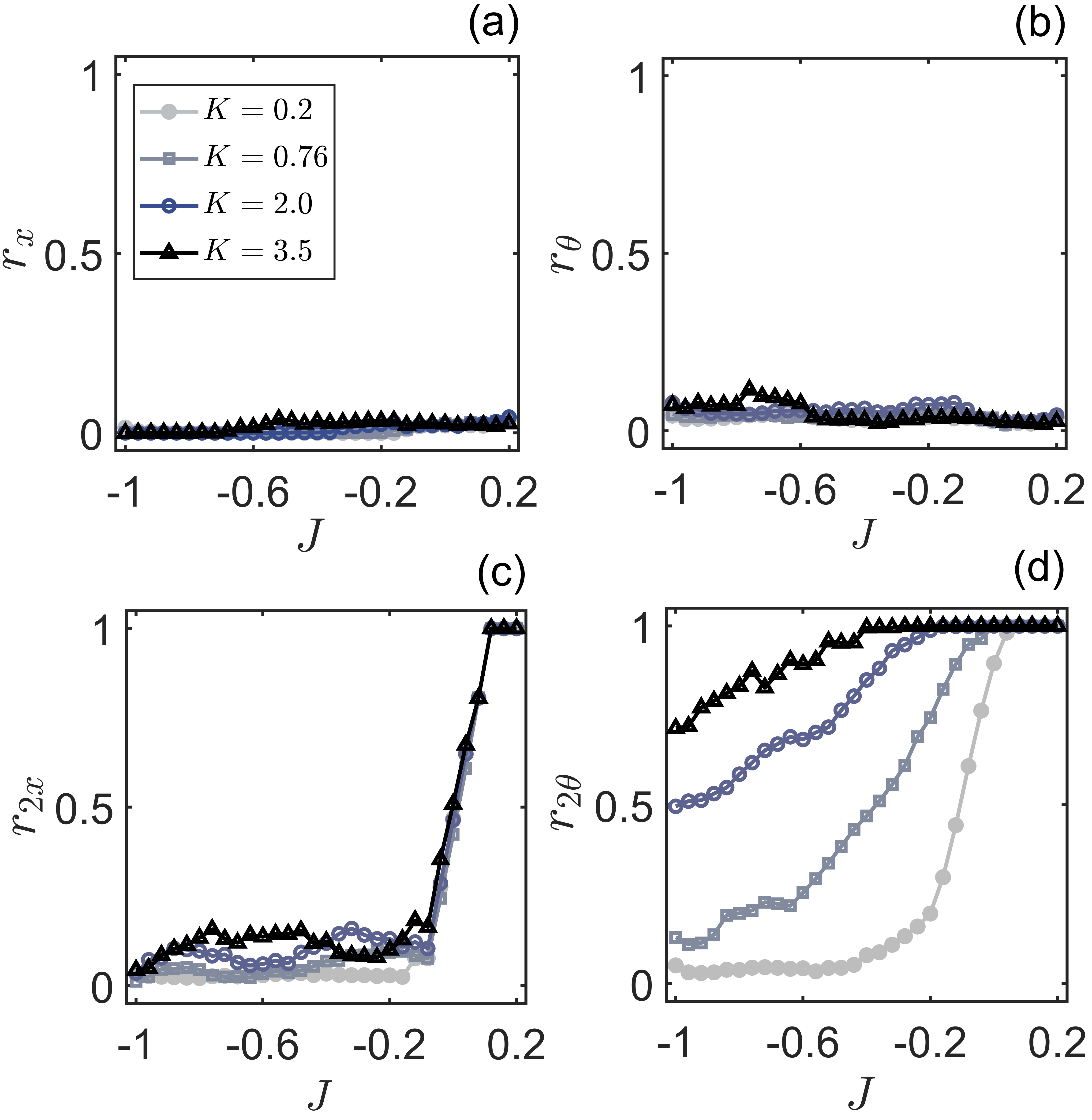}
    \caption{Variation of $r_{x(\theta)}$ and $r_{2x(\theta)}$ with $J$ for distinct values of $K$. We codify the system for $(dt, T, N)=(0.1,5000,10^3)$ and calculate the order parameters after wiping out the first $90\%$ data.}
    \label{mixed state}
\end{figure}

Interestingly, the swarmalators exhibit slight oscillations and eventually shift their positions slowly over time. This slow behavior is somewhat reminiscent of glassy phenomena. Figure~\ref{mixed state velocity} investigates this slow relaxation by plotting the mean velocity $\langle v \rangle$ $(=\sqrt{\Delta x^2+\Delta\theta^2})$ of the system over time. Panels (a) and (b) of Fig.~\ref{mixed state velocity} illustrate that for small system sizes $(N=3,5)$, the system gets gradually cooler and eventually freezes at $10^5$ steps. In contrast, panels (c) and (d) of Fig.~\ref{mixed state velocity} show that for larger system sizes ( $N=50, 200)$, although the system gradually cools as before, it never appears to freeze fully. Instead, finite fluctuations in $\langle v \rangle$ appear to persist indefinitely. We ran simulations for up to $10^7$ steps and never saw the motion die out. To distinguish this state, we examine the order parameters $r_{x(\theta)}$ (see Fig.~\ref{mixed state}(a) and (b)) and $r_{2x(\theta)}$ (notice Fig.~\ref{mixed state}(c) and \ref{mixed state}(d)) over $J$ for various $K$. We observe that all the order parameters behave the same at distinct $K$ except $r_{2\theta}$. For larger values of $K$, $r_{2\theta}$ increases and approaches unity. 

\subsubsection{$\theta$ (antiphase) sync state}

\begin{figure}[hpt]
    \centering
    \includegraphics[width=\linewidth]{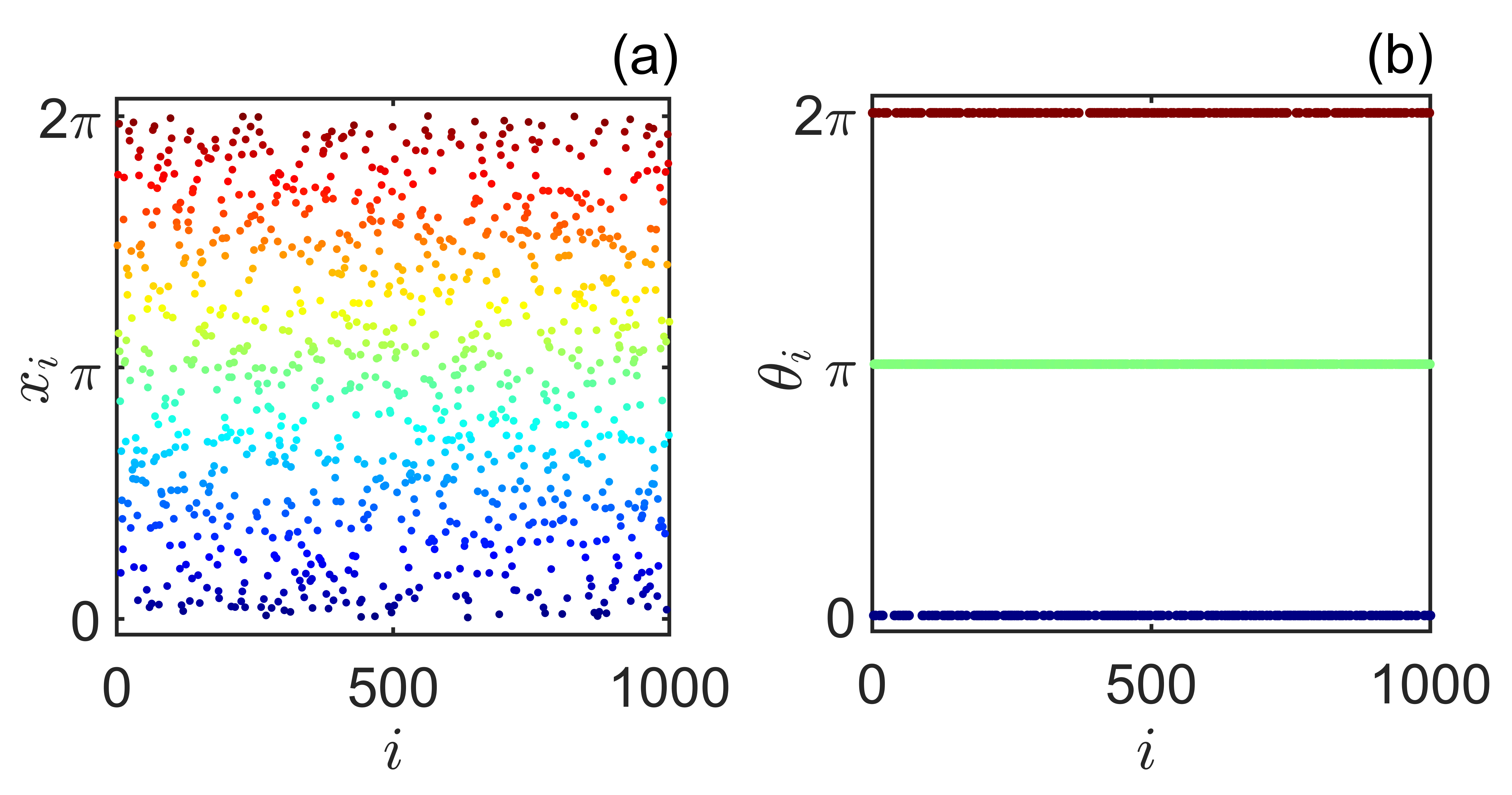}
    \caption{$\theta$ (antiphase) sync state for $(J,K)=(0.0,0.5)$. (a) The position of the swarmalators is plotted (colors representing their positions) against their indices. (b) Scatter plot of the swarmalators' phase concerning their index (colored according to their phases),  Simulation parameters are $(dt, T, N)=(0.1,5000,10^3)$. Initially, both their phases and positions are randomly chosen from $[-\pi,\pi]$. }
    \label{index_theta_antiphase}
\end{figure}

This state arises exclusively for the special case $J=0$, where spatial dynamics are inactive. As a result, the spatial configuration remains as initially set, uniformly random. See Fig.~\ref{index_theta_antiphase}(a), where we plot the positions of the swarmalators according to their indices. Rather, they exhibit a phase separation of $\pi$ (look at Fig.~\ref{index_theta_antiphase}(b)). We can find the stability of this state via standard, finite-$N$ linearization (we take $x_i = \dfrac{2\pi i}{N}$ linearly spaced and $\theta_i = \pi \Theta(i-(\frac{N}{2}+1))$, where $\Theta()$ is the Heaviside function). The eigenvalues for this state are as indicated in Table~\ref{table2}, where $C_i$ are the positive constants that depend on $N$. This corroborates our numerical findings that the state is stable if $K>0$ when it exists for $J=0$.

\begin{table}[h]
    \caption{This table demonstrates the eigenvalues and multiplicity of the Jacobian for the $\theta$ (antiphase) sync state at distinct system sizes.}
    \centering
    \begin{tabular}{ccccc}  \vspace{2pt}
        $N$ value \hspace{1 cm} & (Eigenvalue, Multiplicity)\\ \hline
        \vspace{6pt}
        2 \hspace{1 cm}  &$(0,~2),~ (-K,~2)$\\ 
        \vspace{6pt}
        3 \hspace{1 cm}  &$(0, ~3),~ (-\dfrac{2K}{3}, ~1), (\dfrac{K}{3},~2)$ \\
        \vspace{6pt}
        4 \hspace{1 cm}  &$(0, ~4),~ (-\dfrac{K}{2}, ~4)$ \\
        \vspace{6pt}
        6 \hspace{1 cm} &$(0,~ 6),~ (-\dfrac{2K}{3}, ~2), (-\dfrac{K}{3}, ~4)$ \\
        \vspace{6pt}
        8 \hspace{1 cm} &$(0,~ 8),~ (-\dfrac{K}{4},~ 4),~ (-\dfrac{(1+\sqrt{2})K}{4}, ~4)$ \\
        \vspace{6pt}
        $N$ \hspace{1 cm} &$(0, ~N),~ (- C_i(N) K,~ N)$ \\
        \hline
    \end{tabular}
     \label{table2}
\end{table}

\section{Conclusion} \label{summary}
Our contribution is the first study of swarmalators with Winfree-type coupling, where the elements fire pulses according to a function $P(\theta)$ received by the other elements via a response function $R(\theta)$. This type of coupling is observed in many systems, such as fireflies or neurons, and in some sense generalizes the Kuramoto-type coupling. Table~\ref{table3} shows that Winfree coupling recovers the states seen with Kuramoto coupling (static $\pi$, phase wave, and async states) but also produces three new collective states ($x,\theta$-antiphase sync states and the intermediate mixed state). On the other hand, the active async state that appears with Kuramoto-type coupling was not observed.

We determined the stability thresholds for four of these six states; the stability of the mixed and $x$-antiphase sync state remain out of reach. A natural question for follow-up work is to try and pin these down. Take the antiphase state. As discussed, its properties are somewhat unusual: an infinite number of equilibrium densities $P_{\theta}$ satisfy the state, yet only one, the bell-shaped distribution well fit by a Gaussian, appears to be stable. Why is that? Why is it fit so well by a Gaussian? And is it truly the only stable equilibrium, or could others exist that the numerics failed to reveal?

Future work could also relax some of the idealizations we made in the work. For instance, one could on the natural frequencies $(\omega_i, \nu_i)$ or draw them from some distribution. Heterogeneities in the coupling constants would also be interesting to investigate.

\begin{table}[h]
\caption{Comparison of the emerging states in 1d swarmalators between the Kuramoto coupling and the Winfree coupling:} 
    \centering
    \begin{tabular}{ccccc}  \vspace{2pt}
        State \hspace{0.001 cm} & Kuramoto coupling \hspace{0.01 cm} & Winfree coupling\\ \hline 
        \vspace{6pt}
        Static $\pi$ \hspace{0.01 cm} & \checkmark \hspace{0.01 cm} & \checkmark \\ 
        \vspace{6pt}
        Static async \hspace{0.01 cm}  &\checkmark \hspace{0.01 cm} &\checkmark \\
        \vspace{6pt}
        Static phase wave \hspace{0.01 cm}  &\checkmark \hspace{0.01 cm} &\checkmark \\
        \vspace{6pt}
        Active async \hspace{0.01 cm}  &\checkmark \hspace{0.01 cm} & $\times$ \\
        \vspace{6pt}
        $x$-(antiphase) sync\hspace{0.01 cm}  & $\times$ \hspace{0.01 cm} & \checkmark \\
        \vspace{6pt}
        $\theta$-(antiphase) sync\hspace{0.01 cm}  & $\times$ \hspace{0.01 cm} & \checkmark \\
        \vspace{6pt}
        Intermediate mixed \hspace{0.01 cm}  &$\times$ \hspace{0.01 cm} & \checkmark      \\ \hline
    \end{tabular}
    \label{table3}
\end{table}

\providecommand{\noopsort}[1]{}\providecommand{\singleletter}[1]{#1}%

\end{document}